\documentclass{ws-mpla}

\begin{document}
\markboth{M. HIRAI}
{Determination of $f_0(980)$ structure by fragmentation functions}

\catchline{}{}{}{}{}

\title{DETERMINATION OF $f_0(980)$ STRUCTURE \\
       BY FRAGMENTATION FUNCTIONS
}

\author{M. HIRAI}
\address{Department of Physics, Juntendo University, 
         Inba, Chiba 270-1695, Japan}

\author{S. KUMANO}
\address{Institute of Particle and Nuclear Studies,
          High Energy Accelerator Research Organization (KEK), 
          1-1, Ooho, Tsukuba, Ibaraki, 305-0801, Japan \\
          Department of Particle and Nuclear Studies,
             Graduate University for Advanced Studies \\
           1-1, Ooho, Tsukuba, Ibaraki, 305-0801, Japan}    
 
\author{M. OKA}
\address{Department of Physics, H-27,
             Tokyo Institute of Technology,
             Meguro, Tokyo, 152-8551, Japan}

\author{K. SUDOH}
\address{Institute of Particle and Nuclear Studies,
          High Energy Accelerator Research Organization (KEK), 
          1-1, Ooho, Tsukuba, Ibaraki, 305-0801, Japan}

\maketitle

\pub{Received (Day Month Year)}{Revised (Day Month Year)}

\begin{abstract}
We discuss internal structure of an exotic hadron by using fragmentation
functions. The fragmentation functions for the $f_0(980)$ meson are
obtained by a global analysis of $e^++e^- \rightarrow f_0+X$ data.
Quark configuration of the $f_0(980)$ could be determined 
by peak positions and second moments of the obtained fragmentation
functions.

\keywords{exotic hadron, fragmentation function.}
\end{abstract}

\ccode{PACS Nos.: 12.39.Mk, 13.87.Fh, 13.66.Bc}

\vspace{-0.20cm}
\section{Introduction}	
\vspace{-0.10cm}

Internal quark structure of exotic hadrons is an interesting topic.
Information on the structure would be obtained from parton distribution
functions (PDFs), which are determined by using experimental data
in deeply inelastic scattering (DIS). Stable targets are needed 
for DIS experiments, whereas exotic hadrons decay immediately and
their PDFs could not be measured.

In $e^+e^-$ annihilation experiments, fragmentation functions (FFs)
have been measured for produced several hadrons. 
These functions have information on hadronization, 
so that it is suitable for discussing the internal structure of the hadrons.
The FFs have two types: favored and disfavored functions.
The favored function means a fragmentation from a parton which exists
as a constituent quark in the hadron, and the disfavored one means
a hadronization from a sea quark. 
In general, these functions have characteristic features on their peak
positions at small $Q^2$ ($\sim$1 GeV$^2$).
The favored ones correspond to the valence quark distributions
in the PDFs, so that the peak positions of the favored ones exist
at large $z$, but those of the disfavored ones do at small $z$.
In fact, such a behavior can be seen in the FFs of the charged $K$ meson
(see Fig. 11 in Ref.~\refcite{hkns07}).
Investigating the fragmentation functions, 
we can find the flavor constituent in the exotic hadrons.
The $f_0(980)$ meson is considered as a candidate for an exotic hadron,
which has structure beyond naive $q\bar q$ configurations.
Therefore, we determine the FFs for the $f_0(980)$ by a global analysis
and discuss its possible internal structure.

\vspace{-0.15cm}
\section{Global analysis of fragmentation functions}
\vspace{-0.10cm}

The FFs are associated with the non-perturbative part separated
by the factorization theorem, so that these functions should be
determined by using experimental data. 
The differential cross section, which is normalized by the total
cross section $\sigma_{tot}$, is given for the $e^+e^- \rightarrow hX$ 
process as:
\begin{eqnarray}
F^h(z,Q^2) &=& \frac{1}{\sigma_{tot}} 
               \frac{d\sigma (e^+e^- \rightarrow hX)}{dz} , 
\nonumber \\
	       &=& \sum_i C_i(z,\alpha_s) \otimes D_i^h (z,Q^2) .
\end{eqnarray}
Here, the variable $z$ is defined as the energy ratio of the produced
hadron and a parent parton, and it is given by the hadron energy $E_h$
and the center-of-mass energy $\sqrt{s}$ ($=\sqrt{Q^2}$): 
$z \equiv E_h/(\sqrt{s}/2)$. 
In the theoretical calculation with the perturbative QCD, 
this cross section can be expressed by the sum of convolution integrals 
with the coefficient functions $C_i(z, \alpha_s)$ 
and the fragmentation functions $D_i^h(z,Q^2)$.
Here, $\otimes$ denotes the convolution integral,
$f (z) \otimes g (z) = \int^{1}_{z} dy f(y) g(z/y)/y$,
$\alpha_s$ is the running coupling constant, and
$i$ indicates quark flavor or gluon,
$i=u,\ d,\ s,\ \cdot\cdot\cdot,\ g$.

In this analysis, $Q^2$ dependence of the FFs should be taken into account 
because the experimental data exist at several $Q^2$ points. 
The $Q^2$ dependence is calculated by the DGLAP evolution equations. 
To solve the equations numerically, 
initial functions must be defined at initial $Q^2$ ($\equiv Q^2_0$).
A functional form is somewhat arbitrary, which is a cause of
model dependence in the global analysis. 
We adopt a general functional form which is used in other parametrization
groups: 
\begin{equation}
D_{i}^{f_0} (z,Q_0^2) = N_{i}^{f_0} z^{\alpha_{i}^{f_0}}(1-z)^{\beta_{i}^{f_0}} ,
\end{equation}
where parameter $N_i$ can be rewritten by the second moment 
$M_i (\equiv \int dz z D_i^{f_0} (z)$) 
and the beta function of the parameters $\alpha_i$ and $\beta_i$, 
$N_i=M_i/B(\alpha_i+1, \beta_i+1)$.
The second moments have a physical meaning of energy conservation, 
$\sum_h M_i^h=1$. 
The $M_i$, $\alpha_i$, and $\beta_i$ are free parameters, 
which are optimized by a $\chi^2$ analysis.

As the initial functions of $f_0(980)$, we assume the relation, 
$D_{\bar{q}}^{f_0}(z)=D_{q}^{f_0}(z)$.
%
%
Then, the following five functions 
$D_{u}^{f_0} (z,Q_0^2)$  $(=D_{d}^{f_0} (z,Q_0^2))$, 
$D_{s}^{f_0} (z,Q_0^2)$, $D_{g}^{f_0} (z,Q_0^2)$, and
$D_{c}^{f_0} (z,m_c^2)$, $D_{b}^{f_0} (z,m_b^2)$,
are expressed by a number of parameters.
Here, the functions for up-quark, strange-quark, and gluon
are defined at $Q_0^2$=1 GeV$^2$, and heavy flavor functions
are inserted into the calculation at the mass thresholds,  
$m_c$=1.43 GeV and $m_b$=4.3 GeV.
%
We note that differences between the functions of the up- and
strange-quarks would come from its mass difference.
In principle, a threshold should be also provided for the strange quark
at its mass threshold, $m_s \sim 0.2$ GeV. Since perturbative QCD would
not be applicable at this low scale, which is of the order of $\Lambda_{QCD}$,
the initial scale is defined at $Q_0^2$=1 GeV$^2$ as the lower limit
in the theory. For this reason, the function of the strange-quark 
should differ intrinsically from that of the up-quark at the initial scale.
However, flavor separation cannot be made only by the $e^+e^-$ experimental data. 
We need data from other processes, e.g., semi-inclusive DIS.

\begin{table*}[t]
\tbl{Possible $f_0(980)$ configurations and their relations to
     the second moments and the peak positions for 
     the fragmentation functions of $f_0(980)$.}
{\begin{tabular}{cccc} \toprule 
Type                   & Configuration 
                       & Second moments
                       & Peak positions      \\
\colrule
Strange    $q\bar q$   & $s\bar s$                 
                       & $M_u  <   M_s \lesssim M_g$    
                       & $z_u^{\rm max}<z_s^{\rm max}$   \\  
Tetraquark (or $K\bar K$) & $(u\bar u s\bar s+d\bar d s\bar s)/\sqrt{2}$  
                       & $M_u \sim M_s \lesssim M_g$
                       & $z_u^{\rm max} \sim z_s^{\rm max}$   \\  
\botrule
\end{tabular}}
\end{table*}

\begin{figure}[b]
\vspace{-3mm}
\centerline{\psfig{file=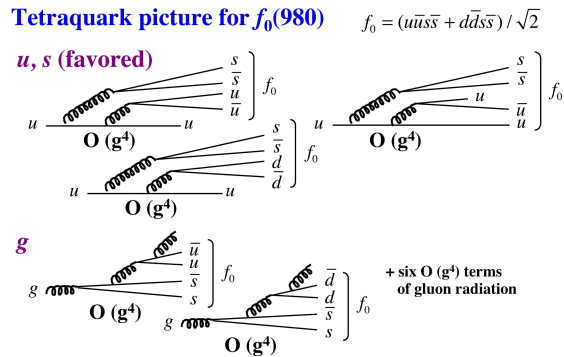,width=75mm}}
\vspace{-0.2cm}
\caption{Schematic diagrams for $f_0$ production in the tetraquark picture.
\protect\label{fig:tetraquark}}
\end{figure}
To discuss the internal structure for $f_0(980)$, 
four configurations are considered, 
non-strangeness $q\bar{q}$, naive $s\bar{s}$, tetraquark, 
and glueball.\cite{f0FFs}
For simplicity, two configurations are listed in Table. 1.
This configuration can be determined 
by the relations in the peak positions and the second moments of the FFs. 
In the tetraquark configuration, for example, 
the functions of the up- and strange-quarks become the favored functions. 
These peak positions are the same: $z_u^{\rm max} \sim z_s^{\rm max}$.
In addition, the relations of the second moments are suggested 
by order counting of the coupling in schematic diagrams.
Hadronizations from the up- and strange-quarks can be expressed 
by the same diagrams as shown in Fig. \ref{fig:tetraquark}. 
Their contributions to the cross sections are reflected
in the second moments, therefore these moments will roughly
become the same: $M_u \sim M_s$. By the same estimation, 
relations for other configurations are obtained
(see Table. 1 in Ref.~\refcite{f0FFs}).

\vspace{-0.20cm}
\section{Results and discussion}
\vspace{-0.10cm}

\begin{figure}[pt]
\centerline{\psfig{file=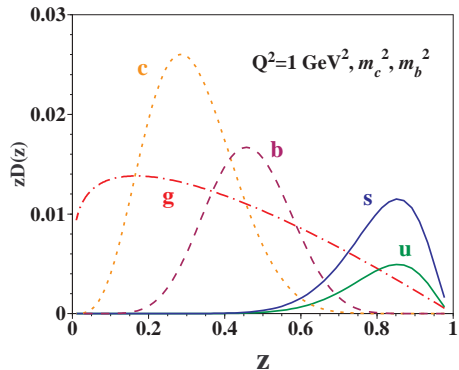,width=60mm}}
\vspace{-0.1cm}
\caption{Obtained fragmentation functions for $f_0(980)$. 
The functions of the up-quark, strange-quark, and gluon are
shown at $Q^2$=1 GeV$^2$, and heavy flavor functions of the charm-
and bottom-quarks are at $Q^2=m_c$ and $m_b$, respectively.
\protect\label{fig:f0-ffs}}
\end{figure}

The fragmentation functions are determined by the global analysis 
with the $f_0(980)$ production data in the $e^+e^-$ annihilation
experiments.\cite{e+e-f0-data} Total $\chi^2$/DOF is 0.907,
which is a reasonable value in the $\chi^2$ analysis.
Obtained FFs are shown in Fig. ~\ref{fig:f0-ffs}.
We find that the peak positions of the functions for
the up- and strange-quarks exist at large $z$.
These are the favored functions, and these peaks are located
at the same points, $z_u^{\rm max} \sim z_s^{\rm max}$.
It indicates the tetraquark configuration as the internal structure
of $f_0(980)$.
%
On the other hand, values of the obtained second moments are 
$ M_u = 0.0012 \pm 0.0107$,
$ M_s = 0.0027 \pm 0.0183$, and
$ M_g = 0.0090 \pm 0.0046$.
The second moment of the up-quark is less than that of
the strange-quark, which indicates the naive $s\bar{s}$ configuration.
This conflicting result is due to low accuracy of the current 
experimental data. In fact, uncertainties of the second moments
are large. At this stage, the internal configuration cannot be
determined clearly. To discuss the quark configuration in details, 
these uncertainties must be reduced by including precise data
in the analysis. Therefore, we hope to get high accuracy data,
which could be provided from the BELLE and BarBar experiments.

\vspace{-0.20cm}
\section*{Acknowledgments}
\vspace{-0.10cm}
The authors were partially supported by the Grant-in-Aid
for Scientific Research from the Japanese Ministry
of Education, Culture, Sports, Science, and Technology. 

\vspace{-0.20cm}


\end{document}